\documentclass[letter]{aa}

\usepackage[dvips]{graphicx}
\usepackage{txfonts}
\usepackage{natbib}        
\bibpunct{(}{)}{;}{a}{}{,} 

\newcommand{\corrections}[1]{#1}

\newcommand{\brG}{Br-\ensuremath{\gamma}}
\newcommand{\photocenter}{\ensuremath{p}}
\newcommand{\photocenterD}{\ensuremath{\vec{p}}}
\newcommand{\phase}{\ensuremath{\phi}}
\newcommand{\PAdisk}{\ensuremath{\mathrm{PA_{disk}}}}
\newcommand{\PAstar}{\ensuremath{\mathrm{PA_{star}}}}
\newcommand{\idisk}{\ensuremath{i_\mathrm{{disk}}}}
\newcommand{\istar}{\ensuremath{i_\mathrm{{star}}}}
\newcommand{\alphaPsa}{\object{Fomalhaut}}

\newcommand{\base}{\ensuremath{\mathrm{B{}}}}
\newcommand{\wave}{\ensuremath{\lambda{}}}

\begin{document}

\title{The spin-orbit alignment of the Fomalhaut planetary system probed by optical long baseline interferometry\ \thanks{Based on observations collected at the VLTI (ESO Paranal, Chile), with the 082.C-0376 program from the AMBER Guaranteed Time of the Osservatorio Astrofisico di Arcetri (INAF, Italy).}}
\titlerunning{The spin-orbit alignment of Fomalhaut probed by OLBIN}

\author{J.-B.~Le~Bouquin\inst{1} \and O.~Absil\inst{2}\thanks{FNRS Postdoctoral Researcher} \and M.~Benisty\inst{3} \and F.~Massi\inst{3} \and A.~M\'erand\inst{1} \and S.~Stefl\inst{1}}

\institute{European Southern Observatory, Casilla 19001, Santiago 19, Chile \and
Institut d'Astrophysique et de G\'eophysique, Universit\'e de Li\`ege, 17 all\'ee du Six Ao\^ut, B-4000 Sart Tilman, Belgium \and
INAF - Osservatorio Astrofisico di Arcetri, Largo E. Fermi 5, 50125 Firenze, Italy}

\offprints{J.B.~Le~Bouquin \\ \email{jlebouqu@eso.org}}

\date{Received 16/02/2009 ; Accepted 09/04/2009}

\abstract
{}
{We discuss the spin-orbit orientation of the Fomalhaut planetary system composed of a central A4V star, a debris disk, and a recently discovered planetary companion.}
{We use spectrally resolved, near-IR long baseline interferometry to obtain precise spectro-astrometric measurements across the \brG{} absorption line. The achieved astrometric accuracy of $\pm{}3\,\mu$as and the spectral resolution $R=1500$ from the AMBER/VLTI instrument allow us to spatially and spectrally resolve the rotating photosphere.}
{We find a position angle $\PAstar=65\degr \pm 3\degr$ for the stellar rotation axis, perfectly perpendicular with the literature measurement for the disk position angle ($\PAdisk=156\fdg0 \pm 0\fdg3$). This is the first time such test can be performed for a debris disk, and in a non-eclipsing system. Additionally, our measurements suggest unexpected backward-scattering properties for the circumstellar dust grains.}
{Our  observations validate  the  standard scenario  for star  and planet formation, in which  the angular momentum of the planetary systems are expected to be collinear with the stellar spins.}  


\maketitle


\section{Introduction}

At 7.7\,pc from the Sun, \alphaPsa{} is one of the closest main sequence star surrounded by a spatially resolved debris disk. It has been studied at many different wavelengths ranging from the visible to the sub-millimetre regime. At long wavelengths (far-infrared and beyond), the disk shows a pair of intensity maxima interpreted as the ansae of an inclined ring about 140\,AU in radius \citep{Holland-2003jan,Stapelfeldt-2004sep}. This particular geometry was subsequently confirmed with HST/ACS imaging \citep{Kalas-2005Natur}, which also provided high accuracy estimations of the disk inclination and position angle: $\idisk=65\fdg9 \pm 0\fdg4$ and $\PAdisk= 156\fdg0 \pm 0\fdg3$, \corrections{under the assumption that the disc is intrinsically circular}. In the scattered light images, dust particles are confined in a narrow ring about 25\,AU in width, with sharp edges that suggest the presence of a planetary body at about 120\,AU \citep{Quillen-2006oct}. It was also noted that the centre of the ring is shifted \corrections{by 15\,AU} with respect to the central star position, pointing on still poorly understood dynamical effects that should also be related to the presence of planetary-mass or sub-stellar companions. Recently, the presence of a planetary companion at the expected orbital distance (119\,AU) was finally confirmed with HST/ACS coronagraphic imaging \citep{Kalas-2008Sci}, which significantly boosted the general interest in this system.

\corrections{The circumstellar dust around stars with ages above $\sim$10~Myr might be partially a remnant from the primordial disc.} The disks could also be replenished by the populations of planetesimals that were not used to build up planet \citep{Mann-2006jun}. These leftovers are supposed to produce dust by mutual collisions or cometary activity. The study of those debris disks provides one of the best means to explore the properties (size, density, orientation) and evolution of planetary systems.

In standard planet formation scenarios, the disk  and  the associated planets are expected to  be in the equatorial plane of the central star, like our Kuiper Belt which is within $2\degr$ of the invariable plane  of the  Solar system \citep{Brown-2004apr}.  Yet, as far as debris disks (and circumstellar disks in general) are concerned, this simple observational prediction has never been put into test.  More  generally, the common  assumption   of  collinearity  between planetary orbital  axis and  stellar spins has only been tested  on the Solar system, with  a relevant accuracy ($\sim 1\degr$),  and on a few transiting  extrasolar  giant planets  with  a  rather poor  accuracy \citep[$\sim10\degr$, by the mean of the Rossiter-McLaughlin spectroscopic  effect,][]{Rossiter-1924jul}.  In at least  one  case, XO-3b, a large spin-orbit  misalignment   is  suggested   by  the observations \citep[$70\degr \pm 15\degr$,][]{Hebrard-2008sep}.  Additionally,  the recent  controversy about  the  spin-orbit alignment  of  HD~17156b  illustrates well  the difficulty of such a technique, with measured spin-orbit misalignments of $62\degr \pm 25\degr$ by \citet{Narita-2008apr}  and later $9\fdg4 \pm 9\fdg3$  by \citet{Cochran-2008aug}.  Up to now,  this test has never been performed in non-transiting systems, because it is generally impossible to recover the orientation of the rotational axis of the central star.  

With the advent of spectrally dispersed optical long baseline interferometry, it becomes possible to put this hypothesis into test by spatially and spectrally resolving the stellar photospheres, even marginally. Indeed, this technique allows the orientation of stellar axes to be quickly (geometrically) recovered by differential astrometry across a photospheric line enlarged by the stellar rotation. The purpose of this Letter is to perform this test in the case of \alphaPsa{}, which advantageously combines a fast projected rotational velocity of $93\,$km/s \citep{Royer-2007feb} and a relatively large apparent diameter of $\Phi=2.2\,$mas \citep{DiFolco-2004nov}.

\begin{figure}[t]
    \centering
    \includegraphics[scale=0.5]{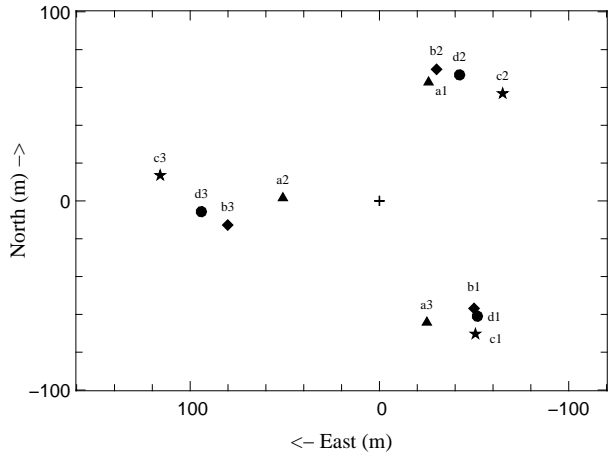}
    \caption{\corrections{Projected baselines} of the  observations. Each observation (a, b,
      c  and  d, draw with different symbols)  is  composed  of  3  baselines  (1,2,3) along which the projections of the 2D
      spectro-astrometric displacements are measured
      (see Fig.~\ref{fig:shifts}).} 
    \label{fig:uv}
\end{figure}


\section{Observation and data reduction} \label{sec:observations}

Data have  been collected at  the Very Large  Telescope Interferometer \citep[VLTI,][]{Haguenauer-2008spie}  \corrections{with  the  AMBER beam combiner} \citep{Petrov-2007mar} recording  spectrally dispersed fringes \corrections{between $1.92-2.28\,\mu$m} at  a spectral resolution of  $R=1500$. The Auxiliary Telescopes   (ATs)   were    placed   at   stations   D0--H0--G1   and A0--K0--G1.  Observations were assisted  by the  fringe-tracker FINITO, allowing an  individual exposure time of 200\,ms  in AMBER \citep{LeBouquin-2008spie_b}. \corrections{The seeing ranged $0.8"$ to $1.2"$.}
FINITO provided     good     locking     ratio    and     fringe     tracking performance. Observations  of calibration stars  were interleaved with those  of the  science target.  Unlike most  classical interferometric observations,  the only prerequisite  for calibrators  is not  to show spectro-astrometric  features  around the  \brG{}  line. We  therefore favour cool giants because they  have less pronounced \brG{} lines and smaller  rotation   velocities.  The  observation  log   is  given  in Table~\ref{tab:obs_log}.
\corrections{Regarding the magnitudes, \alphaPsa{} is $1.00\,\mathrm{K^{mag}}$, $\gamma$~Scl is $1.72\,\mathrm{K^{mag}}$, and 88~Aqr is $0.98\,\mathrm{K^{mag}}$. Because it is fainter, data on $\gamma$Scl have been obtained with slightly longer integration time. This has no incidence on the differential analysis made in this paper.}

\begin{table}[t]
  \centering
  \caption{Observation log with the number of files (\#), the number of frames per file (NDIT) and the elementary integration time (DIT). Calibration stars are in italic.}
  \begin{tabular}[c]{lccccccc} \hline \hline \vspace{-0.25cm}\\
    Date      & Baseline & Target & \# & NDITxDIT
    \vspace{0.03cm}\\ \hline \vspace{-0.30cm} \\
    Obs. a) &&&& Total: 600s \\
    2008-12-05T00:20 & D0-H0-G1 & Fomalhaut   & 5 & 200x0.2s \\
    2008-12-05T01:13 & D0-H0-G1 &  88Aqr       & 5 & 200x0.2s \\
    2008-12-05T01:30 & D0-H0-G1 & Fomalhaut   & 5 & 200x0.2s \\
    2008-12-05T01:48 & D0-H0-G1 &  $\gamma$Scl      & 5 & 150x0.5s \\
    2008-12-05T02:06 & D0-H0-G1 & Fomalhaut   & 5 & 200x0.2s \\
    2008-12-05T02:24 & D0-H0-G1 &  88Aqr       & 5 & 200x0.2s
    \vspace{0.03cm}\\ \hline \vspace{-0.30cm} \\
    Obs. b) &&&& Total: 200s \\
    2008-12-08T02:15 & A0-K0-G1 & Fomalhaut   & 5 & 200x0.2s \\ 
    2008-12-08T02:31 & A0-K0-G1 &  88Aqr       & 5 & 200x0.2s
    \vspace{0.03cm}\\ \hline \vspace{-0.30cm} \\
    Obs. c) &&&& Total: 200s \\
    2008-12-09T00:10 & A0-K0-G1 & Fomalhaut   & 5 & 200x0.2s \\ 
    2008-12-09T00:27 & A0-K0-G1 &  88Aqr       &     5 & 200x0.2s
    \vspace{0.03cm}\\ \hline \vspace{-0.30cm} \\
    Obs. d) &&&& Total: 200s \\
    2008-12-09T01:31 & A0-K0-G1 & Fomalhaut   & 5 & 200x0.2s \\ 
    2008-12-09T01:51 & A0-K0-G1 & $\gamma$Scl      & 5 & 120x0.5s \\\hline
  \end{tabular}
  \flushleft
  \label{tab:obs_log}
\end{table}

Phases of interferometric fringes were computed using the latest version of the \texttt{amdlib} package (version 2.99, Chelli et al., private communication) and the interface provided by  the  Jean-Marie  Mariotti  Center\footnote{\texttt{http://www.jmmc.fr/data\_processing\_amber.htm}}.  Following  a  standard  fringe selection criterion,  we have kept  the 80\% best  frames. Consecutive observations of  \alphaPsa{} were grouped into single  data points with enhanced signal-to-noise  ratio (SNR), referred to  as observations a, b, c, and d in Table~\ref{tab:obs_log}.  
Accordingly,  the  observations of  the  calibration  stars were  also appended into single, high quality measurements of the instrumental phase. \corrections{Such an average introduces baseline smearing. Considering the worst case of 2h, we estimated the smearing of the astrometric signal to be less than 20\%. This is a reasonable price to pay to obtain clear detections on all dataset.} After the calibration of scientific data, we  therefore end up with a set of 4  observations of  Fomalhaut (see  Table~\ref{tab:obs_log}),  each of them composed of the  3 phase spectra from the AMBER spectrograph.  While signal was also detected in the visibility, we decided  to focus  the analysis  of  this letter on the phase, which contains the  astrometric quantity. 

\begin{figure}[t]
    \centering
    \includegraphics[scale=0.5]{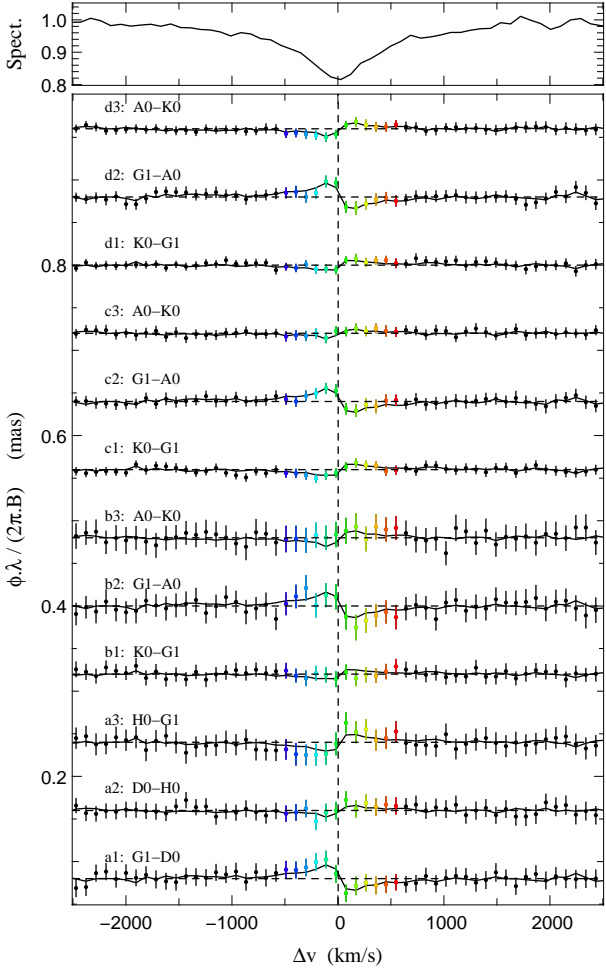}
    \caption{Top: continuum-normalized spectrum obtained with AMBER for a spectral window centered on the \brG{} line. Bottom: Differential phase measurements converted into spectro-astrometric shifts $\photocenter(\wave)$ in mas. Colours represent the position within the line, from blue to red. \corrections{The solid lines link the results, reprojected on the baselines, of the global 2D astrometric solution $\photocenterD(\wave)$ performed independently for each spectral channel. Phase spectra are shifted vertically by $0.08\,$mas.}}
    \label{fig:shifts}
\end{figure}

\corrections{The unavoidable atmospheric dispersion appears as a global  curvature of  the fringe phases over the $K$ band. It cancels perfectly when computing the sum of the 3 phase spectra, demonstrating that the baselines are consistent to each others in term of sign (so in direction on sky). Subtracting the  instrumental phase from the  scientific phase removed all fine  spectral features down  to a level of  $0\fdg5$ (uncertainty level).  However, atmospheric  dispersion was not  perfectly canceled out. It was efficiently removed  by fitting the spectral range $2.03-2.27\,\mu$m with an \emph{ad-hoc} polynomial law of order 3}. As a matter of  fact, phase measurements are relative to  the continuum value  which we  suppose to  have  a zero  phase. It  may not  be perfectly true because  of the gravity darkening in  the fast rotating atmosphere. However, such effect would not bias our measurement.

The  calibrated  differential  phases  \phase{}  were  converted  into differential  astrometric shifts  \photocenter{} using  the well-known formula   for   marginally   resolved   interferometric   observation \citep{Lachaume-2003mar}:
\begin{equation}
\photocenter = -\frac{\phase}{2\pi} \cdot \frac{\wave}{\base}
\end{equation}\vspace{0.07cm}
\noindent\base{} is the length of the interferometric baseline and \wave{} is the effective wavelength of the spectral channel considered. \photocenter{} represents the projection, on the baseline direction, of the 2D   photocenter  in   the  plane   of  the   sky,   hereafter  called \photocenterD{}. A single astrometric solution (i.e. a single 2D vector \photocenterD{})  was fitted  to all  observations available  within a single  spectral channel (i.e.  the 12  projections \photocenter{}).  This  global fit  is perfectly linear  and has  the  advantage of  showing  the ultimate  astrometric performance. However, the resulting astrometric signal $\photocenterD(\wave)$  can   be  slightly  biased   (towards  reduced amplitude) because of the smearing that may occur in case of different spectral calibration  between the observations. This has  no impact on the astrophysical interpretation made in the following.  \corrections{The differential astrometric  shifts $\photocenter(\wave)$ are represented  by the dots and the error bars in Fig.~\ref{fig:shifts}. The result of  the global fit by the  2D vector  $\photocenterD(\wave)$ is represented, reprojected on the baselines, by the solid lines that connect the fitted values.} 

Uncertainties were propagated into the astrometric vector \photocenterD{}  by standard formulae. We did not take into account the uncertainty on the baseline length nor on the spectral calibration because they affect all spectral bins in the same way and therefore have negligible effect on our final observable. The astrometric error ellipses were found to be almost circular thanks to the relatively uniform $u,v$ plane coverage provided by our observations (see Fig.~\ref{fig:uv}). Uncertainties are in the range $\pm2\,\mu$as to $\pm6\,\mu$as\corrections{, with typically $\pm3\,\mu$as across the \brG{} line.} It is among the most precise spectro-astrometric measurements ever achieved, if not the best.

When talking about  potential bias, our main concern  was the presence of a  possible artifact on the  differential phase $\phase(\lambda){}$ due to the particular spectrum  of our target (A4V, i.e. strong \brG{} absorption  line)  in comparison to  the  calibrator  spectrum  (K  giants, i.e. almost no \brG{}  line). However, an instrumental artifact linked to  the   AMBER  spectrograph  would  \emph{rotate}   on  sky  between observations  a),  b),  c)  and  d), following  the  rotation  of  the associated   baseline  triangles   (see   table~\ref{tab:obs_log}  and fig.~\ref{fig:uv}).  The good  agreement between  the  signal obtained during  3  different   nights  and  on  different  configurations confirms its astrophysical origin.  As an additional test, we observed  other   fast   rotating   stars   and  obtained   various   kinds   of spectro-astrometric  signals,  as well  as  expected non-detection  on small photospheres.


\section{The Fomalhaut rotation axis} \label{sec:rotation_axis}

\begin{figure*}[t]
    \centering
    \includegraphics[scale=0.5]{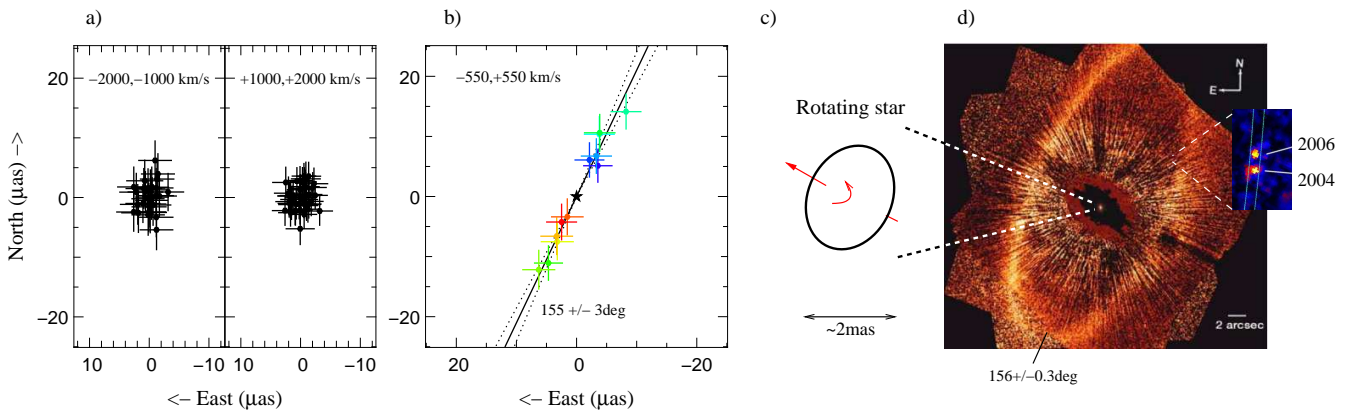}
    \caption{ AMBER spectro-astrometric positions $\photocenterD(\wave)$ in the continuum (a) and across the Br-$\gamma{}$ absorption  line (b). Colors refer to the wavelength bin,  as shown in  Fig.~\ref{fig:shifts}. The signature of  the  rotating  photosphere  (c)  is clearly  detected  and  is compared to the debris disk and the planetary companion (d) imaged in the  visible by  \citet{Kalas-2008Sci}. For the  sake of clarity, the  astrometric error ellipses are  represented by their projection into the North and East directions.} 
    \label{fig:result}
\end{figure*}

Our observations represent the most precise spectro-astrometric displacements ever measured with the AMBER instrument ($\pm3\,\mu$as in the \brG{} line). To extract the scientifically useful information, we plot the spectro-astrometric photocenters onto the plane of the sky. As a first check, we plot the photocenters in two spectral windows of continuum around the \brG{} line (Fig.~\ref{fig:result}a). The dispersion of the data points as a function of wavelength in the continuum is fully compatible with our estimated uncertainties ($\pm3\,\mu$as) and shows no systematic behaviour or unexpected features at our precision level.

On  the  contrary,  a  significant  astrometric  displacement  is detected   when   plotting  the   signal   across   the  \brG{}   line (Fig.~\ref{fig:result}b).  The photocentre is  displaced in one direction on the  blue wing of the line,  and is identically displaced on the  opposite direction  on the red  wing. The shift  is compatible with zero in the centre of  the line, and its maximum amplitude in the line wings reaches about $15\,\mu$as. This is a clear detection of the effect  of rotation on the stellar  photosphere, from which the \brG{} absorption line originates. Interpreting such a small astrometric  shift does not  require deep modelling of  the fast rotating atmosphere.   In particular, the  astrometric displacement is necessarily  perpendicular  to  the  position  angle  of  the  stellar rotation axis.   Fitting our  complete data set,  we find  the stellar rotation  axis   to  have  position  angle  $\PAstar   =  65\degr  \pm 3\degr$.  This  value is  perfectly  perpendicular  to the  literature measurement for the disk angle $\PAdisk=156\fdg0 \pm 0\fdg3$ (Fig.~\ref{fig:result}c and \ref{fig:result}d).  

Theoretically, an estimation of the inclination angle \istar{} is also possible  from  spatially   and  spectrally  resolved  interferometric observations   by   using   rapidly  rotating   stellar   photospheric models \citep{Domiciano-2002oct}. However,  the limited accuracy and spectral  resolution of our data does  not allow  the stellar inclination  to be constrained  in a meaningful way.  We note  that \alphaPsa{} has been observed intensively   in  broad-band   interferometry with the VINCI/VLTI instrument    \citep{DiFolco-2004nov,LeBouquin-2006may},   but   these observations, even when  considered  as  a whole,  did  not allow  to recover the inclination  angle, mostly due to the  very small expected oblateness ratio  of the \alphaPsa{}  photosphere ($\sim$2\%, O.~Absil et al.,  in preparation). Weak constraints on  the stellar inclination can nevertheless  be extracted  from simple physical  assumptions. The maximum  permissible break-up  velocity  for \alphaPsa{}  is close  to $390\,$km/s according to von~Zeipel  equations with typical values for A5V stars ($2\,$M$_\odot$  and $1.7\,$R$_\odot$). Assuming a projected rotational  velocity of $93\,$km/s  \citep{Royer-2007feb}, it  gives a lower  limit $\istar\,$$>$$\,15\degr$.  We conclude  that  the current existing  constraints are  compatible  with the  literature value  for the disk inclination ($\idisk=65.9\degr$), but does not represent a significant test.  Interferometric observations at higher precision, higher  spectral resolution,  and  ideally with  longer baselines  may certainly answer this important point.  

Besides constraining the orientation of the stellar photosphere, spatially and spectrally resolved observations also constrain the direction of its spin vector. Taking into account that the \brG{} line is in absorption, we conclude that the South-East part of the star is moving towards us. If we assume that the orbital angular momentum of the planetary companion points towards the same direction as the stellar spin, the Western side of the debris disk is located on the observer's side of the sky plane. Combined with the  observed increased brightness in the Eastern part of the disk \citep{Kalas-2005Natur}, it suggests that the dust grains are mostly backward-scattering. This finding is in contradiction with the well-known forward-scattering properties of circumstellar dust grains in our Solar system, generally assumed to be true in all debris disks \citep[e.g.,][]{Weinberger-1999nov,Kalas-2005Natur}. Unfortunately, this question only appeared as a very interesting by-product at the time of the study, and our observing strategy has not been specifically designed to answer it. Especially, no check star is available in the data-set to perfectly secure the sign of the AMBER phase.

\corrections{ We did our best to calibrate the phase sign a-posteriori. First we checked that a positive delay corresponds to a negative phase as measured by the AMBER reduction package we used. Secondly, we converted the phase curvature across the K-band (due to atmospheric refraction) into actual position in sky. We found that the blue part of the band is indeed shifted toward the zenith (as it should be). However, even if we are confident, we cannot draw definite conclusions before a real spectro-astrometric reference has been observed. }


\section{Conclusions} \label{sec:discussion}

From  the technical  point of  view, these  observations are  a strong validation of  the remarkable potential  of the spectro-interferometer AMBER/VLTI, especially  in its  Medium Resolution mode  and associated with  the   fringe-tracker  FINITO.  Stacking   17\,min  of  effective integration on a $\mathrm{K}=0.94^\mathrm{mag}$  star, we were able to reach  a spectro-astrometric  accuracy of  $\pm{}3\,\mu$as. Associated with a  spectral resolution $R=1500$,  this was enough to  resolve the rotating  photosphere of  \alphaPsa{},  with an  apparent diameter  of $2.2\,$mas and a rotational velocity of $93\,$km/s. We believe this to be the first  detection of the astrometric displacement  created by an absorption line in a fast rotating photosphere. 

Fitting our complete dataset, we find the position angle of \alphaPsa{} rotation axis to be $\PAstar=65\degr \pm 3\degr$, perfectly perpendicular to the literature measurement for the disk angle $\PAdisk=156\fdg0 \pm 0\fdg3$. This is the first time such a test could be performed outside the Solar system for a non-eclipsing system. However, as when using the Rossiter-McLaughlin method, the inclination angle remains unknown. We can only conclude that there is strong evidence, but no definite proof, that the planetary system of \alphaPsa{} is in the equatorial plane of the central star. Additionally, by determining the direction of the stellar spin vector, we demonstrate that the technique is able to constrain the scattering properties of the dust grains surrounding \alphaPsa{}. \corrections{Such measurements as presented in this paper are} only possible thanks to the remarkable combination of a fast stellar rotation, a large stellar apparent diameter, the presence of a resolved disk, and of a planetary companion with resolved orbital motion.

We plan to continue observing the \alphaPsa{} system to provide even better constraints on the rotation of the central star, as well as to extend the study to other fast rotating stars hosting resolved debris disks.

\begin{acknowledgements}
JBLB warmly thanks the complete VLTI team of the Paranal Observatory. This work has made use of the Smithsonian/NASA Astrophysics Data System (ADS) and of the Centre de Donnees astronomiques de Strasbourg (CDS). Graphics were performed with \texttt{Yorick}\footnote{\texttt{http://yorick.sourceforge.net}}.
\end{acknowledgements}


\begin{thebibliography}{20}
\expandafter\ifx\csname natexlab\endcsname\relax\def\natexlab#1{#1}\fi

\bibitem[{{Brown} \& {Pan}(2004)}]{Brown-2004apr}
{Brown}, M.~E. \& {Pan}, M. 2004, \aj, 127, 2418

\bibitem[{{Cochran} {et~al.}(2008){Cochran}, {Redfield}, {Endl}, \&
  {Cochran}}]{Cochran-2008aug}
{Cochran}, W.~D., {Redfield}, S., {Endl}, M., \& {Cochran}, A.~L. 2008, \apjl,
  683, L59

\bibitem[{{Di Folco} {et~al.}(2004){Di Folco}, {Th{\'e}venin}, {Kervella},
  {Domiciano de Souza}, {Coud{\'e} du Foresto}, {S{\'e}gransan}, \&
  {Morel}}]{DiFolco-2004nov}
{Di Folco}, E., {Th{\'e}venin}, F., {Kervella}, P., {et~al.} 2004, \aap, 426,
  601

\bibitem[{{Domiciano de Souza} {et~al.}(2002){Domiciano de Souza}, {Vakili},
  {Jankov}, {Janot-Pacheco}, \& {Abe}}]{Domiciano-2002oct}
{Domiciano de Souza}, A., {Vakili}, F., {Jankov}, S., {Janot-Pacheco}, E., \&
  {Abe}, L. 2002, \aap, 393, 345

\bibitem[{{Haguenauer} {et~al.}(2008){Haguenauer}, {Abuter}, {Alonso},
  {Argomedo}, {Bauvir}, {Blanchard}, {Bonnet}, {Brillant}, {Cantzler}, {Derie},
  {Delplancke}, {Di Lieto}, {Dupuy}, {Durand}, {Gitton}, {Gilli}, {Glindemann},
  {Guniat}, {Guisard}, {Haddad}, {Hudepohl}, {Hummel}, {Jesuran}, {Kaufer},
  {Koehler}, {Le Bouquin}, {L{\'e}v''que}, {Lidman}, {Mardones}, {M{\'e}nardi},
  {Morel}, {Percheron}, {Petr-Gotzens}, {Phan Duc}, {Puech}, {Ramirez},
  {Rantakyr{\"o}}, {Richichi}, {Rivinius}, {Sahlmann}, {Sandrock},
  {Sch{\"o}ller}, {Schuhler}, {Somboli}, {Stefl}, {Tapia}, {Van Belle},
  {Wallander}, {Wehner}, \& {Wittkowski}}]{Haguenauer-2008spie}
{Haguenauer}, P., {Abuter}, R., {Alonso}, J., {et~al.} 2008, in SPIE Conf.,
Vol. 7013

\bibitem[{{H{\'e}brard} {et~al.}(2008){H{\'e}brard}, {Bouchy}, {Pont},
  {Loeillet}, {Rabus}, {Bonfils}, {Moutou}, {Boisse}, {Delfosse}, {Desort},
  {Eggenberger}, {Ehrenreich}, {Forveille}, {Lagrange}, {Lovis}, {Mayor},
  {Pepe}, {Perrier}, {Queloz}, {Santos}, {S{\'e}gransan}, {Udry}, \&
  {Vidal-Madjar}}]{Hebrard-2008sep}
{H{\'e}brard}, G., {Bouchy}, F., {Pont}, F., {et~al.} 2008, \aap, 488, 763

\bibitem[{{Holland} {et~al.}(2003){Holland}, {Greaves}, {Dent}, {Wyatt},
  {Zuckerman}, {Webb}, {McCarthy}, {Coulson}, {Robson}, \&
  {Gear}}]{Holland-2003jan}
{Holland}, W.~S., {Greaves}, J.~S., {Dent}, W.~R.~F., {et~al.} 2003, \apj, 582,
  1141

\bibitem[{{Kalas} {et~al.}(2008){Kalas}, {Graham}, {Chiang}, {Fitzgerald},
  {Clampin}, {Kite}, {Stapelfeldt}, {Marois}, \& {Krist}}]{Kalas-2008Sci}
{Kalas}, P., {Graham}, J.~R., {Chiang}, E., {et~al.} 2008, Science, 322, 1345

\bibitem[{{Kalas} {et~al.}(2005){Kalas}, {Graham}, \&
  {Clampin}}]{Kalas-2005Natur}
{Kalas}, P., {Graham}, J.~R., \& {Clampin}, M. 2005, \nat, 435, 1067

\bibitem[{{Lachaume}(2003)}]{Lachaume-2003mar}
{Lachaume}, R. 2003, \aap, 400, 795

\bibitem[{{Le Bouquin} {et~al.}(2008){Le Bouquin}, {Abuter}, {Bauvir},
  {Bonnet}, {Haguenauer}, {di Lieto}, {Menardi}, {Morel}, {Rantakyr{\"o}},
  {Schoeller}, {Wallander}, \& {Wehner}}]{LeBouquin-2008spie_b}
{Le Bouquin}, J.-B., {Abuter}, R., {Bauvir}, B., {et~al.} 2008, in SPIE Conf.,
Vol. 7013

\bibitem[{{Le Bouquin} {et~al.}(2006){Le Bouquin}, {Labeye}, {Malbet}, {Jocou},
  {Zabihian}, {Rousselet-Perraut}, {Berger}, {Delboulb{\'e}}, {Kern},
  {Glindemann}, \& {Sch{\"o}ller}}]{LeBouquin-2006may}
{Le Bouquin}, J.-B., {Labeye}, P., {Malbet}, F., {et~al.} 2006, \aap, 450, 1259

\bibitem[{{Mann} {et~al.}(2006){Mann}, {K{\"o}hler}, {Kimura}, {Cechowski}, \&
  {Minato}}]{Mann-2006jun}
{Mann}, I., {K{\"o}hler}, M., {Kimura}, H., {Cechowski}, A., \& {Minato}, T.
  2006, \aapr, 13, 159

\bibitem[{{Narita} {et~al.}(2008){Narita}, {Sato}, {Ohshima}, \&
  {Winn}}]{Narita-2008apr}
{Narita}, N., {Sato}, B., {Ohshima}, O., \& {Winn}, J.~N. 2008, \pasj, 60, L1+

\bibitem[{{Petrov} {et~al.}(2007){Petrov}, {Malbet}, {Weigelt}, {Antonelli},
  {Beckmann}, {Bresson}, {Chelli}, {Dugu{\'e}}, {Duvert}, {Gennari},
  {Gl{\"u}ck}, {Kern}, {Lagarde}, {Le Coarer}, {Lisi}, {Millour}, {Perraut},
  {Puget}, {Rantakyr{\"o}}, {Robbe-Dubois}, {Roussel}, {Salinari}, {Tatulli},
  {Zins}, {Accardo}, {Acke}, {Agabi}, {Altariba}, {Arezki}, {Aristidi},
  {Baffa}, {Behrend}, {Bl{\"o}cker}, {Bonhomme}, {Busoni}, {Cassaing},
  {Clausse}, {Colin}, {Connot}, {Delboulb{\'e}}, {Domiciano de Souza},
  {Driebe}, {Feautrier}, {Ferruzzi}, {Forveille}, {Fossat}, {Foy},
  {Fraix-Burnet}, {Gallardo}, {Giani}, {Gil}, {Glentzlin}, {Heiden},
  {Heininger}, {Hernandez Utrera}, {Hofmann}, {Kamm}, {Kiekebusch}, {Kraus},
  {Le Contel}, {Le Contel}, {Lesourd}, {Lopez}, {Lopez}, {Magnard}, {Marconi},
  {Mars}, {Martinot-Lagarde}, {Mathias}, {M{\`e}ge}, {Monin}, {Mouillet},
  {Mourard}, {Nussbaum}, {Ohnaka}, {Pacheco}, {Perrier}, {Rabbia}, {Rebattu},
  {Reynaud}, {Richichi}, {Robini}, {Sacchettini}, {Schertl}, {Sch{\"o}ller},
  {Solscheid}, {Spang}, {Stee}, {Stefanini}, {Tallon}, {Tallon-Bosc}, {Tasso},
  {Testi}, {Vakili}, {von der L{\"u}he}, {Valtier}, {Vannier}, \&
  {Ventura}}]{Petrov-2007mar}
{Petrov}, R.~G., {Malbet}, F., {Weigelt}, G., {et~al.} 2007, \aap, 464, 1

\bibitem[{{Quillen}(2006)}]{Quillen-2006oct}
{Quillen}, A.~C. 2006, \mnras, 372, L14

\bibitem[{{Rossiter}(1924)}]{Rossiter-1924jul}
{Rossiter}, R.~A. 1924, \apj, 60, 15

\bibitem[{{Royer} {et~al.}(2007){Royer}, {Zorec}, \&
  {G{\'o}mez}}]{Royer-2007feb}
{Royer}, F., {Zorec}, J., \& {G{\'o}mez}, A.~E. 2007, \aap, 463, 671

\bibitem[{{Stapelfeldt} {et~al.}(2004){Stapelfeldt}, {Holmes}, {Chen}, {Rieke},
  {Su}, {Hines}, {Werner}, {Beichman}, {Jura}, {Padgett}, {Stansberry},
  {Bendo}, {Cadien}, {Marengo}, {Thompson}, {Velusamy}, {Backus}, {Blaylock},
  {Egami}, {Engelbracht}, {Frayer}, {Gordon}, {Keene}, {Latter}, {Megeath},
  {Misselt}, {Morrison}, {Muzerolle}, {Noriega-Crespo}, {Van Cleve}, \&
  {Young}}]{Stapelfeldt-2004sep}
{Stapelfeldt}, K.~R., {Holmes}, E.~K., {Chen}, C., {et~al.} 2004, \apjs, 154,
  458

\bibitem[{{Weinberger} {et~al.}(1999){Weinberger}, {Becklin}, {Schneider},
  {Smith}, {Lowrance}, {Silverstone}, {Zuckerman}, \&
  {Terrile}}]{Weinberger-1999nov}
{Weinberger}, A.~J., {Becklin}, E.~E., {Schneider}, G., {et~al.} 1999, \apjl,
  525, L53

\end{thebibliography}


\end{document}